\documentclass{article}

\usepackage{PRIMEarxiv}

\usepackage[utf8]{inputenc} 
\usepackage[T1]{fontenc}    
\usepackage{hyperref}       
\usepackage{url}            
\usepackage{booktabs}       
\usepackage{amsfonts}       
\usepackage{nicefrac}       
\usepackage{microtype}      
\usepackage{lipsum}
\usepackage{fancyhdr}       
\usepackage{graphicx}       
\graphicspath{{media/}}     
\usepackage{amsmath,amsthm,amsfonts,amssymb,bm}

\title{An Improved Structured Mesh Generation Method Based on Physics-informed Neural Networks
}

\author{
  Xinhai Chen, \quad Jie Liu, \quad Junjun Yan, \quad Zhichao Wang,\quad Chunye Gong\\
  Laboratory of Software Engineering for Complex System
 \\
  National University of Defense Technology \\
  Changsha, China\\
  \texttt{\{chenxinhai16, liujie\}@nudt.edu.cn} \\
}

\begin{document}
\maketitle

\begin{abstract}
Mesh generation remains a key technology in many areas where numerical simulations are required. As numerical algorithms become more efficient and computers become more powerful, the percentage of time devoted to mesh generation becomes higher. In this paper, we present an improved structured mesh generation method. The method formulates the meshing problem as a global optimization problem related to a physics-informed neural network. The mesh is obtained by intelligently solving the physical boundary-constrained partial differential equations. To improve the prediction accuracy of the neural network, we also introduce a novel auxiliary line strategy and an efficient network model during meshing. The strategy first employs a priori auxiliary lines to provide ground truth data and then uses these data to construct a loss term to better constrain the convergence of the subsequent training. The experimental results indicate that the proposed method is effective and robust. It can accurately approximate the mapping (transformation) from the computational domain to the physical domain and enable fast high-quality structured mesh generation.
\end{abstract}


\section{Introduction}
Numerical simulation has proved extremely useful for the design and analysis procedures in the fields of scientific research and engineering technology \cite{Speziale1998,Sarun2003}. Many numerical schemes, such as the finite element method (FEM), finite volume method (FVM), and finite difference method (FDM), require the discretization of the geometric (physical) domain \cite{Shu2003,Fries2010,Strouboulis2001}. The discretization procedure, also known as mesh generation (meshing), is a prerequisite for numerical solving. Since the quality of the generated mesh has a significant impact on the accuracy and the efficiency of simulations, the study of mesh generation techniques has received a great deal of attention and has become the centerpiece of computer-aided design and engineering \cite{Liseikin2007,Karman2006,Lowrie2011}.

Structured meshes are composed of uniformly shaped elements with no irregular nodes in their interior. The regular connectivity of structured meshes brings many advantages for numerical simulation. For example, structured meshes offer higher numerical accuracy, less cell count than unstructured meshes, and more straightforward implementation of higher order numerical schemes \cite{Ronald2015,Subramanian1996}. They also produce sparse banded system matrices and are ideal for multigrid acceleration and parallel computing \cite{Huang2017}. With the development of computing power and the increasing complexity of the physical problem, mesh generation has become one of the main performance bottlenecks in the whole numerical simulation process, especially structured mesh generation. Thus, developing an efficient mesh generation technique with high-quality structured meshing capability is desired.

In recent years, deep neural networks (DNNs) have been used with great success in the field of numerical simulations \cite{Chen1995,Pang2021,Chen2021b}. They offer an end-to-end surrogate model utilizing the composition of various neural structures and activation functions. The well-trained network can be applied to accurately predict high-resolution fields \cite{Octavi2020}, aerodynamic performance \cite{Yao2018}, mesh quality \cite{Chen2020,Chen2021}, and flow vortex \cite{Deng2022}. Regarding the intelligent solution of partial differential equations (PDEs), Raissi et al. \cite{Raissi2019,Raissi2020} proposed a novel framework, physics-informed neural networks (PINNs), to learn the physical conservation laws inherent in PDEs. The framework embeds the governing equations and initial/boundary conditions into the loss function of the neural network and employs optimizers to guide the gradient descent direction. After suitable training, the network model is able to provide a nonlinear function approximator for the underlying PDE systems. PINN and its variants have been widely used to learn the quantities of interest from flow visualization or solve different forms of equations, including N-S equations, Maxwell’s equations, and Schrödinger’s equations \cite{Kharazmi2020,Fang2021,Cai2022}. The universal approximation properties of PINNs provide new avenues for structured mesh generation.

Chen et al. \cite{Chen2022} first applied neural networks to automatic mesh generation and introduced a differential mesh generation method MGNet based on unsupervised neural networks. The main insight of the MGNet is its simplicity and its computation speed. Specifically, the method employs a neural network to study the intrinsic mapping relationships (transformation) between computational and physical domains. During the training, the governing Laplace equations, as well as the boundary curves of the input geometry, are embedded in the loss function as penalty terms to guide the gradient descent optimization. Their results showed that the trained MGNet is able to achieve fast meshing using feedforward prediction techniques and generate high-quality structured meshes. But despite its effectiveness, there are still some obvious improvements to be made, such as more refined construction of loss function and exploitation of the input geometry. In other words, the rigid connectivity (meshing rules) is too restrictive during the meshing process, and a few mispredicted nodes can dramatically disrupt the mesh quality (discussed in Section 4). Moreover, incorporating more valid a priori knowledge can help better mitigate the overall distortion of the mesh. Addressing these limitations and improving the usefulness of the neural network-based mesh generation method is the objective of this work.

In this paper, we present an improved structured mesh generation method. The method formulates the generation task as an optimization problem related to a physics-informed neural network. The optimization process is constrained by intelligently solving a governing PDE within the given geometric domain. In this process, an auxiliary line strategy is employed to offer a priori knowledge for network training. We sample input point data from the auxiliary line and feed them into the construed neural network. These sampled points are embedded in the loss function as measured data (ground truth) and serve as a data-driven term to continuously calibrate the convergence in each training epoch. The experimental results on different examples prove that the proposed method is effective and robust. It can estimate the cell overlap problem in the neural network-based generators and enable fast high-quality structured mesh generation. Currently, the method is implemented for two-dimensional problems. The extension to three-dimensional cases is being actively pursued.

The rest of the paper is organized as follows. In Section 2, we first provide a recap of traditional structured mesh generation methods and the philosophy of physics-informed neural networks. In Section 3, we present the implementation details of the improved neural network-based structured mesh generation method. The proposed method is then applied to different mesh generation tasks, and the performance of the method is shown in Section 4. Finally, we conclude the paper and discuss the future works in Section 5.

\section{A Short Description of the Structured Mesh Generation and PINNs}
\label{sec:headings}

\subsection{Structured Mesh Generation}
A structured mesh is formed by intersections of the two coordinate lines (for two-dimensional cases) of a curvilinear coordinate system. There exists a regular connection relationship between the mesh points. This relationship is usually represented using the matrix notation $(i, j)$, where $i$ and $j$ are indices of the two curvilinear coordinates. Due to the regularity, each non-boundary point has the same number of neighbors, and neighboring points in the physical domain (determined by a prescribed set of geometric boundaries) are adjoining in the curvilinear coordinate system. One way of viewing structured mesh generation is the procedure of mapping a regular Cartesian mesh in the coordinate system (usually called the computational domain), via a transformation, to the physical domain to provide a mesh consisting of regular cells. Since the 1970s, the automatic generation of structured mesh has received a lot of interest \cite{Subramanian1996,Ronald2015,Zhang2012}. 

Algebraic methods and PDE methods are the two most commonly used structured mesh generation methods. Algebraic methods use algebraic interpolation to describe the potential mapping relationship between the computational domain $(\xi, \eta)$ and the physical domain $(x, y)$. One of the most important categories is transfinite interpolation, which is initially designed in \cite{Gordon1973}. This interpolation provides a set of algebraic equations to discretize any quadrilateral domain using its boundary parametrization. The general form of the algebraic method is formulated as:

\begin{equation}
\begin{gathered}
\vec{r}(\xi, \eta)=(1-\xi) \vec{r}_l(\eta)+\xi \vec{r}_{\mathrm{r}}(\eta)+(1-\eta) \vec{r}_{\mathrm{b}}(\xi)+\vec{r}_{\mathrm{t}}(\xi) \\
-(1-\xi)(1-\eta) \vec{r}_{\mathrm{b}}(0)-(1-\xi) \vec{r}_{\mathrm{t}}(0) \\
-\xi(1-\eta) \vec{r}_{\mathrm{b}}(1)-\xi \eta \vec{r}_{\mathrm{t}}(1)
\end{gathered}
\end{equation}
where $\vec{r}_l$, $\vec{r}_{\mathrm{r}}$, $\vec{r}_{\mathrm{t}}$, $\vec{r}_{\mathrm{b}}$ denote the left, right, upper and bottom boundaries of the computational domain, respectively. The main advantages of the algebraic method are simplicity and the ease of controlling the shape and density of the mesh cells. However, the drawback is that in the case of deformed geometric boundaries, the traditional algebraic method tends to produce poor-quality cells (e.g., distorted or even overlapped), which weakens its usefulness in complex scenarios.

The PDE method is developed for generating structured meshes under complex geometric boundaries. The core of this method is to obtain the mapping $(\xi, \eta) \longrightarrow(x, y)$ by numerically solving partial differential equations. Based on the governing equation, PDE methods can be subdivided into three main categories: elliptic, parabolic and hyperbolic, among which the most commonly used are the elliptic Poisson and Laplace equation-based mesh generation methods \cite{Thompson1974,Thompson1982}.

Given the boundary curves of the input geometry, the elliptic PDE-based method treats the meshing process as a class of initial boundary value problems and solves the partial differential equation system inside the domain. The governing partial differential equation is of the form
\begin{equation}
\nabla^2 \xi^i=P^i \quad (i=1,2)
\end{equation}
or
\begin{equation}
\left\{\begin{array}{l}
\xi_{x x}+\xi_{y y}=P(\xi, \eta) \\
\eta_{x x}+\eta_{y y}=Q(\xi, \eta)
\end{array}\right.
\end{equation}
where $\nabla$ is Laplace operator, $P^i$, $P$, $Q$ are source terms.

Due to the natural smoothness of the elliptic equations, the elliptic PDE-based mesh generation method can suppress the boundary singularities and keep the gradient discontinuity from propagating into the interior field, thus generating a mesh with good orthogonality. One of the primary bottlenecks of the PDE method is the required computational and meshing overhead, especially for high-resolution or large-scale simulations. Therefore, developing a fast structured mesh generation technique with high-quality meshing capability is desired.

\subsubsection{Physics-informed Neural Network}
With the development of artificial intelligence theory and technology, integrating neural networks into traditional numerical simulations has received much research interest. Recently, pioneering works began to explore the possibility of applying deep neural networks to solve PDEs. Physics-informed neural networks (PINNs) were first introduced in \cite{Raissi2019,Raissi2020} to infer PDE solutions by means of the universal approximation theorem \cite{Chen1995}. For a PDE system of the general form:
\begin{equation}
\begin{aligned}
&\boldsymbol{u}_t+\mathcal{N}_{\boldsymbol{x}}[\boldsymbol{u}]=0, \boldsymbol{x} \in \Omega, t \in[0, T] \\
&\boldsymbol{u}(\boldsymbol{x}, 0)=h(\boldsymbol{x}), \boldsymbol{x} \in \Omega \\
&\boldsymbol{u}(\boldsymbol{x}, t)=g(\boldsymbol{x}, t), t \in[0, T], \boldsymbol{x} \in \partial \Omega
\end{aligned}
\end{equation}
where the spatial domain $\Omega \in \mathbf{R}^d$, $\partial \Omega$ is the boundary of $\Omega$, and $N_x$ is a differential operator.

\begin{figure}[htbp]
  \centerline{\includegraphics[width=0.6\textwidth]{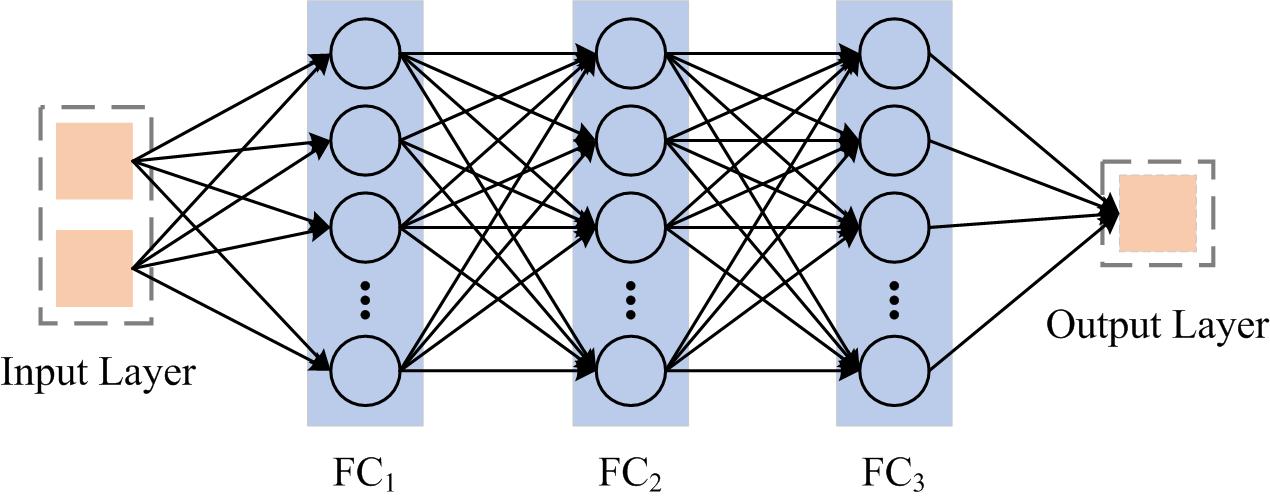}}
  \caption{An example of a physics-informed neural network with three fully connected (FC) layers.}
  \label{fig:fig1}
\end{figure}

PINNs utilize multiple layers of neural units $f_\theta(x, t)$ to automatically approximate the latent solution $\boldsymbol{u}(\boldsymbol{x}, t)$ from high-dimensional parameter spaces. In PINNs, the neurons are fully connected. Figure \ref{fig:fig1} shows an example of a physics-informed neural network with three fully connected layers. During training, the governing equations, as well as the initial/boundary conditions, are embedded in the loss function as penalty terms. The loss function in PINN is defined as:
\begin{equation}
\mathcal{L} o s s=\sum_{i=1}^3 \sum_{j=1}^N\left|e_i\left(\boldsymbol{x}^j, t^j\right)\right|^2
\end{equation}
where
\begin{equation}
\begin{aligned}
&\boldsymbol{e}_1=\frac{\partial}{\partial t} f_\theta(\boldsymbol{x}, t)+\mathcal{N}_{\boldsymbol{x}}\left[f_\theta(\boldsymbol{x}, t)\right] \\
&\boldsymbol{e}_2=\boldsymbol{u}(\boldsymbol{x}, 0)-h(\boldsymbol{x}) \\
&\boldsymbol{e}_3=\boldsymbol{u}(\boldsymbol{x}, t)-g(\boldsymbol{x}, t)
\end{aligned}
\end{equation}

Subsequently, optimization algorithms, such as stochastic gradient descent or quasi-Newton methods, are used to minimize the loss function and update the adjustable variables (weights and biases) in the network model. After suitable training, the trained PINN can work as a function approximator that naturally encodes the underlying physical conservation laws and provide the predictive solutions to partial differential equations.

\section{An Improved Structured Mesh Generation Method Based on Physics-informed Neural Networks}
\subsection{Auxiliary Line Strategy}
In this section, we present an improved structured mesh generation method based on physics-informed neural networks. Given the boundary of a two-dimensional region defined by a series of vertices, we first fit the boundary curves using these given control points. Similar to our previous work \cite{Chen2022}, we use a decision tree regression (DTR)-based regression model \cite{Rokach2005} to approximate the mapping $(\xi, \eta) \longrightarrow(x, y)$ on the boundary curves. The obtained fitting functions are employed to provide sufficient boundary point samples for the subsequent training.

To some extent, it is feasible to constrain the meshing process using the boundary functions combined with governing equations. However, in cases where complex geometries or deformed boundary curvature exist, the above knowledge may not guarantee high-quality mesh generation or requires a large number of training epochs to find an acceptable suboptimal solution. This is because the rigid connectivity (meshing rules) is too restrictive during the meshing process, and a few mispredicted nodes can dramatically disrupt the quality of the mesh. Thus, the next step of the proposed method is to introduce an auxiliary line strategy to improve the usefulness of the neural network-based mesh generation method.

The main idea of the strategy is to offer a priori knowledge for network training. This object is achieved by drawing auxiliary lines in the physical domain. Since the control points on these auxiliary lines are pre-known, we can easily sample a large amount of measured data from the lines. These obtained data can be used as ground truth to enforce constraints on the local mesh point distribution, thus mitigating the overall distortion of the mesh as much as possible.

\begin{figure}[htbp]
  \centerline{\includegraphics[width=0.66\textwidth]{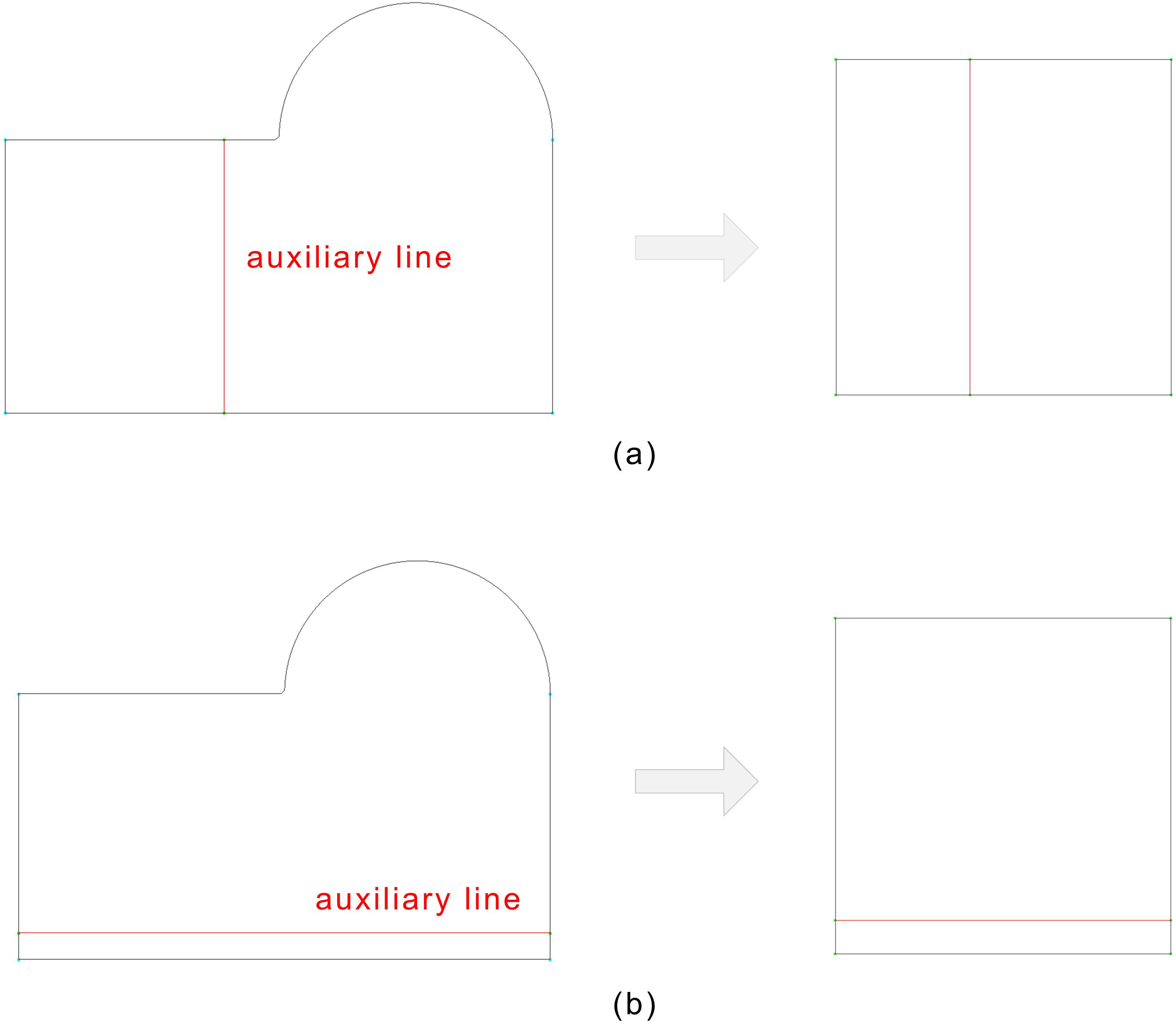}}
  \caption{An example of the introduced auxiliary line strategy on a two-dimensional region. (a) We use a line parallel to the left boundary in the physical domain as an auxiliary line, whose corresponding position in the computational domain is shown on the right. (b) The line parallel to the bottom boundary is selected as the auxiliary line.}
  \label{fig:fig2}
\end{figure}

Figure \ref{fig:fig2} shows an example of the introduced auxiliary line strategy on a two-dimensional region. As a curve line object is difficult to handle, we can simply replace it by drawing a line parallel to the geometric boundary for our purpose. The auxiliary line provides sampled points as measured data (ground truth) in the loss function and formulates a data-driven term to continuously calibrate the convergence in each training epoch. It is worth noting that the number of auxiliary lines used in the mesh generation process is unlimited, and we can employ multiple auxiliary lines to jointly constrain the meshing procedure. Moreover, benefiting from the engineering experience, we can also create auxiliary lines in a more refined way, such as using the medial axis or customized curves for a specific scene. These refined lines can better help the neural network converge to an acceptable local optimum, although some manual interaction will be introduced.

\subsection{Loss Function Construction}
We consider the mesh generation task as an optimization problem. Once the boundary fitting function and auxiliary lines are determined, we can construct a loss function to guide the optimization process. The governing equations used in this paper to control mesh generation are elliptic partial differential equations, which are defined as:
\begin{equation}
\begin{array}{cl}
\alpha x_{\xi \xi}-2 \beta x_{\xi \eta}+\gamma x_{\eta \eta}=0, & (\xi, \eta) \in \Omega \\
\alpha y_{\xi \xi}-2 \beta y_{\xi \eta}+\gamma y_{\eta \eta}=0, & (\xi, \eta) \in \Omega \\
x=f i t_x(\xi, \eta), & (\xi, \eta) \in \partial \Omega \\
y=f i t_y(\xi, \eta), & (\xi, \eta) \in \partial \Omega
\end{array}
\end{equation}
where
\begin{equation}
\begin{aligned}
\alpha &=x_\eta^2+y_\eta^2 \\
\beta &=x_{\xi} x_\eta+y_{\xi} x_\eta \\
\gamma &=x_{\xi}^2+y_{\xi}^2
\end{aligned}
\end{equation}

To this end, a composite loss function that includes the governing equation term, boundary condition term, and auxiliary line term is formulated as follows:
\begin{equation}
Loss=\mathcal{L}_{\text {eqns }}+\lambda_1 \mathcal{L}_{b c s}+\lambda_2 \mathcal{L}_{\text {data }}
\end{equation}
where
\begin{equation}
\mathcal{L}_{\text {data }}=\sum_{j=1}^{N_{\text {data }}}\left(\left|x_{\text {data }}\left(\xi^j, \eta^j\right)-x_{\text {pred }}\left(\xi^j, \eta^j\right)\right|^2+\left|y_{\text {data }}\left(\xi^j, \eta^j\right)-y_{\text {pred }}\left(\xi^j, \eta^j\right)\right|^2\right)
\end{equation}
\begin{equation}
\mathcal{L}_{eqns}=\sum_{i=1}^2 \sum_{j=1}^{N_{\text {eqns }}}\left|e_i\left(\xi^j, \eta^j\right)\right|^2
\end{equation}
\begin{equation}
\mathcal{L}_{bcs}=\sum_{i=3}^4 \sum_{j=1}^{N_{b c s}}\left|e_i\left(\xi^j, \eta^j\right)\right|^2
\end{equation}
\begin{equation}
\begin{aligned}
&\boldsymbol{e}_1=\alpha x_{\xi \xi}-2 \beta x_{\xi \eta}+\gamma x_{\eta \eta} \\
&\boldsymbol{e}_2=\alpha y_{\xi \xi}-2 \beta y_{\xi \eta}+\gamma y_{\eta \eta} \\
&\boldsymbol{e}_3=x-f i t_x(\xi, \eta)\\
&\boldsymbol{e}_4=y-f i t_y(\xi, \eta)
\end{aligned}
\end{equation}

Here, $\mathcal{L}_{eqns}$ and $\mathcal{L}_{bcs}$ denote the residual of the governing equation and boundary conditions, respectively. $\mathcal{L}_{data}$  denotes the loss between the predicted data $x_{pred}$ ($y_{pred}$) and measured data $x_{data}$ ($y_{pred}$) obtained from the auxiliary line. The parameter $N$ represents the number of points randomly sampled in the computational domain. $\lambda_1$ and $\lambda_2$ are coefficients used to overcome the imbalance contribution of different loss terms. 

\subsection{Network Architecture}
\begin{figure}[htbp]
  \centerline{\includegraphics[width=1\textwidth]{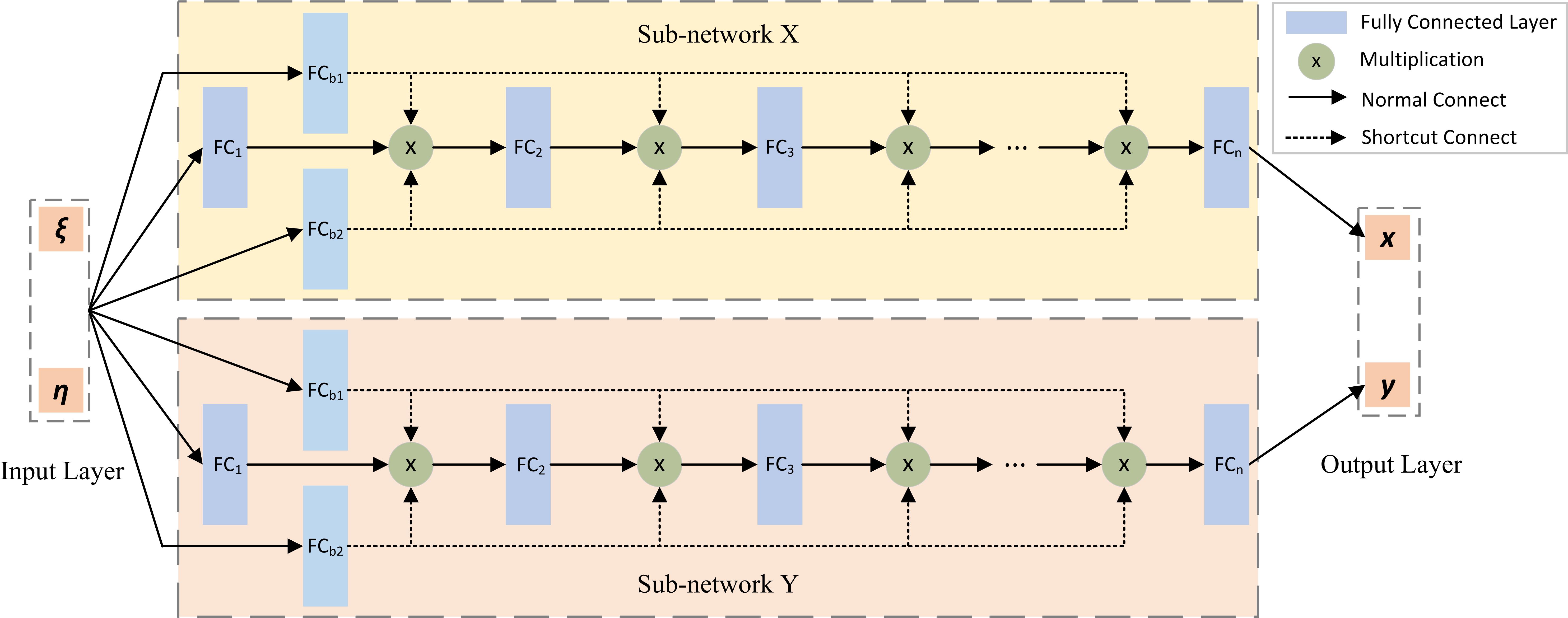}}
  \caption{The network architecture of the proposed method. The network consists of two sub-networks. Each sub-network takes the computational domain coordinates $(\xi, \eta)$ as input and outputs one coordinate dimension ($x$ or $y$) in the physical domain. Two shortcut-based blocks are introduced in each sub-network to enhance the hidden states with residual connections.}
  \label{fig:fig3}
\end{figure}

At completion of the above two steps, we now introduce the network architecture of the proposed method. As can be seen in Figure \ref{fig:fig3}, the network consists of two sub-networks. Each sub-network takes the computational domain coordinates $(\xi, \eta)$ as input and outputs one coordinate dimension ($x$ or $y$) in the physical domain.

Inspired by shortcut and attention-based components widely used for computer vision tasks \cite{Wang2020,Chen2021A}, we introduce two shortcut-based blocks in each sub-network to enhance the hidden states with residual connections. These connections first project the input coordinates to higher dimensional feature space, and secondly employ a point-wise multiplication operation to weight the output of each fully connected layer. The affine transformation in each sub-network is computed as:

\begin{equation}
\begin{aligned}
&F C_{b 1}=\sigma\left(W_{b 1} \cdot \boldsymbol{x}+b_{b 1}\right) \\
&F C_{b 2}=\sigma\left(W_{b 2} \cdot \boldsymbol{x}+b_{b 2}\right) \\
&F C_1=\sigma\left(W_1 \cdot \boldsymbol{x}+b_1\right) \\
&F C_k^{\prime}=\sigma\left(W_k \cdot F C_k+b_k\right), \quad k=1, \ldots, L \\
&F C_{k+1}=\left(1-F C_k^{\prime}\right) \times F C_{b 1}+F C_k^{\prime} \times FC_{b 2}, \quad k=1, \ldots, L \\
&f_\theta=W \cdot F C_{k+1}+b
\end{aligned}
\end{equation}

where $\boldsymbol{x}$ denotes input coordinates $(\xi, \eta)$, the operation $\times$ denotes point-wise multiplication, and $\sigma$ is the activation function. 

Overall, we develop an improved structured mesh generation method. The solution to this problem can be considered as a global optimization problem related to a physics-informed neural network. This process consists of minimizing a loss function defined in Eq. 9 and updating the adjustable variables (weights and biases) to reach a local optimal.

\section{Results and Discussions}
In this section, we perform a series of experiments on the proposed method and compare it with existing neural network-based and traditional mesh generation methods. 

In terms of network size, we do not consider very deep architectures. The neural network used in this work consists of four hidden layers with 30 neurons per layer. Benefiting from this lightweight architecture, we can efficiently conduct the training on the CPUs, which suit well the practical mesh generation environment. For all test cases, we train the network on Intel Intel(R) Xeon 2660 CPUs with the TensorFlow deep learning framework \cite{Abadi2016}. The first- and second-order derivative in loss function is estimated using $tf.gradients()$ in TensorFlow 1.14 based on the chain rule and automatic differentiation.

Activation functions, including $sigmoid$, $swish$, $relu$, $tanh$, play a vital role in neural network training \cite{Ramachandran2017}. These functions introduce nonlinear transformation in each hidden layer, making it possible for neurons to approximate complex mapping relationships. The activation function used in the proposed method is defined as:
\begin{equation}
\sigma(x)=\frac{\sinh x}{\cosh x}=\frac{e^x-e^{-x}}{e^x+e^{-x}}
\end{equation}

\begin{figure}[htbp]
  \centerline{\includegraphics[width=0.78\textwidth]{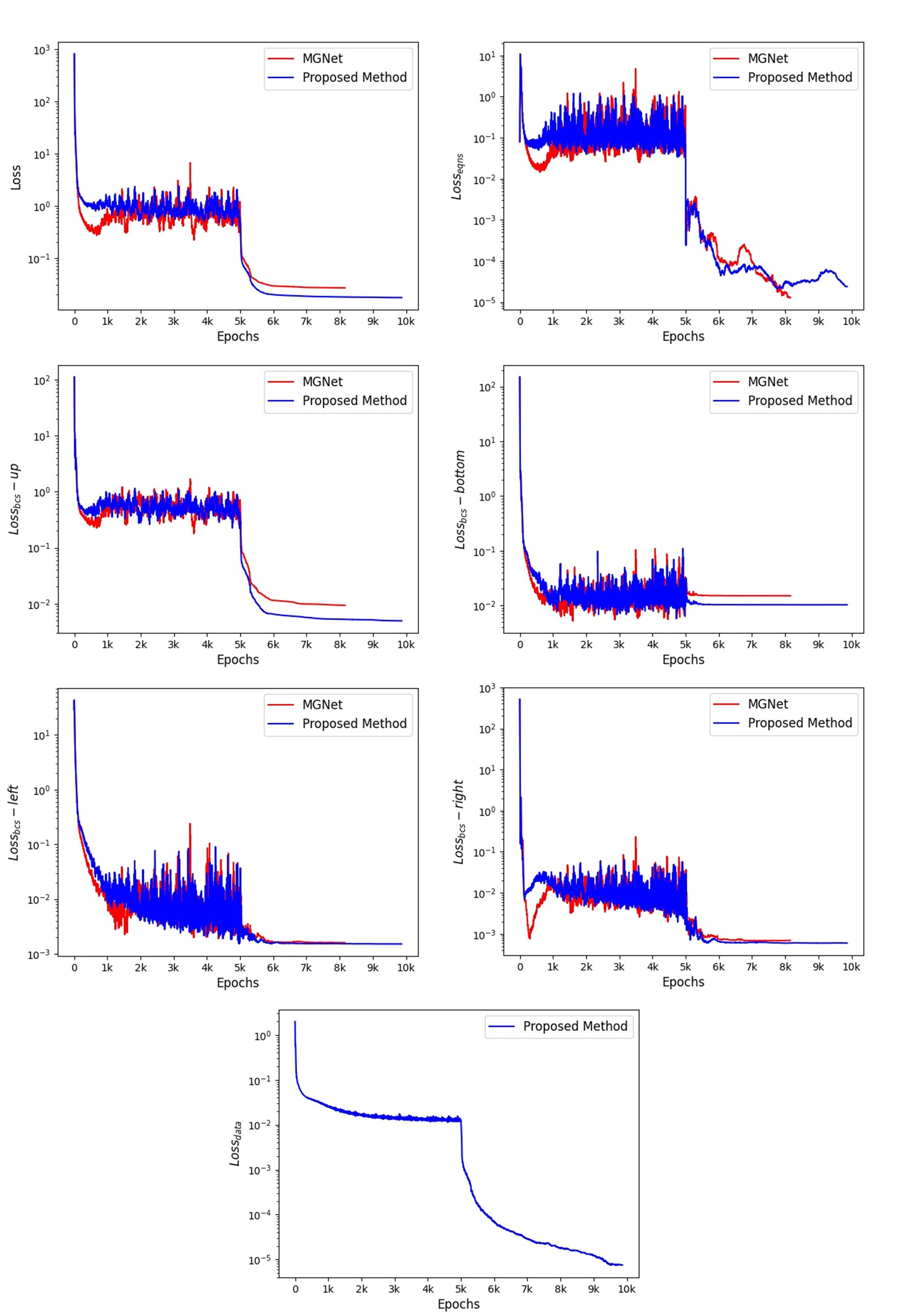}}
  \caption{A comparison of the convergence of two neural network-based generators.}
  \label{fig:fig4}
\end{figure}

For non-convex optimization, we first use the Adam optimizer \cite{Kingma2014} with an initial learning rate of 0.001 to minimize the loss function. The learning rate decays 0.9 every 1000 epochs. Then, we employ a limited-memory quasi-Newton optimizer, L-BFGS-B \cite{Morales2011}, to finetune the network variables and avoid local minima. During Adam-based training, the number of points fed into the neural network is 100 per epoch (batch size), and the total training epoch is 5000. Since the L-BFGS-B optimizer is a full batch approach, we set the training batch to 1000. All the input point samples are randomly extracted from the computational domain.

Inspired by \cite{Wang2020}, we also introduce a dynamic weighting approach to determine the value of the penalizing coefficient $\lambda_1$. This approach is able to adaptively adjust the contribution of different loss terms, thus effectively mitigating the unbalanced gradient pathology in physics-informed neural networks. As for another penalizing coefficient $\lambda_2$, we use a static weighting approach, and the value is fixed to 10.

In the first test case, we employ a 2-D domain to investigate the meshing capability of the neural network-based mesh generation method. Figure \ref{fig:fig4} depicts the convergence of each loss term. To prove the effectiveness of the method, we compare it with an existing neural network-based generator, MGNet \cite{Chen2022}. From the variation curves of each loss value, we can observe that the two-stage (Adam and L-BFGS-B) optimization process is effective in minimizing the loss function. During the Adam phase, the loss value decreases with the increase in the training epoch. After the first 5000 Adam epochs, the loss function converges rapidly under the L-BFGS-B optimizer, and the final outputs are the local optimal solution. We can also see that, in all terms, the proposed method exhibits better convergence results than MGNet. Taking the loss term at the upper boundary $Loss_{bcs}$-$up$ as an example, MGNet gives a relatively low performance (9.370522e-03) in this case, while the proposed method yields 4.939974e-03. Finally, the proposed method outperforms MGNet and achieves a composite loss value of 1.742072e-02.

\begin{figure}[htbp]
  \centerline{\includegraphics[width=1\textwidth]{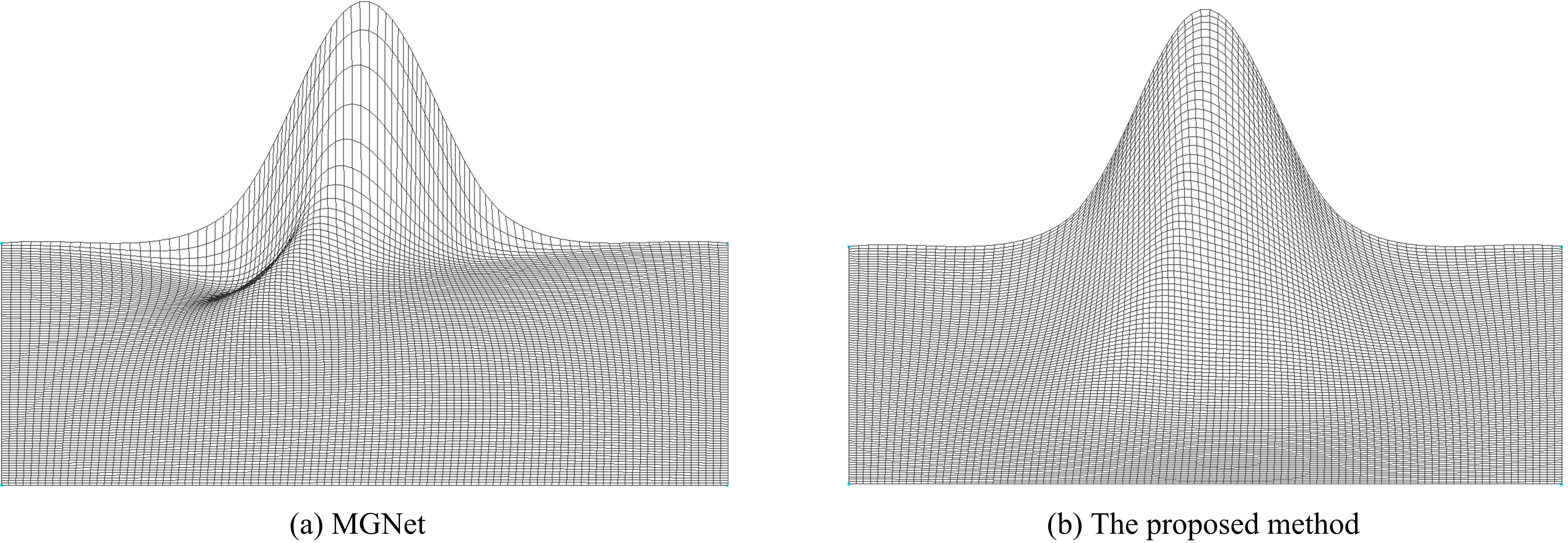}}
  \caption{Visualization results of different neural network-based mesh generation methods.}
  \label{fig:fig5}
\end{figure}

\begin{figure}[htbp]
  \centerline{\includegraphics[width=1\textwidth]{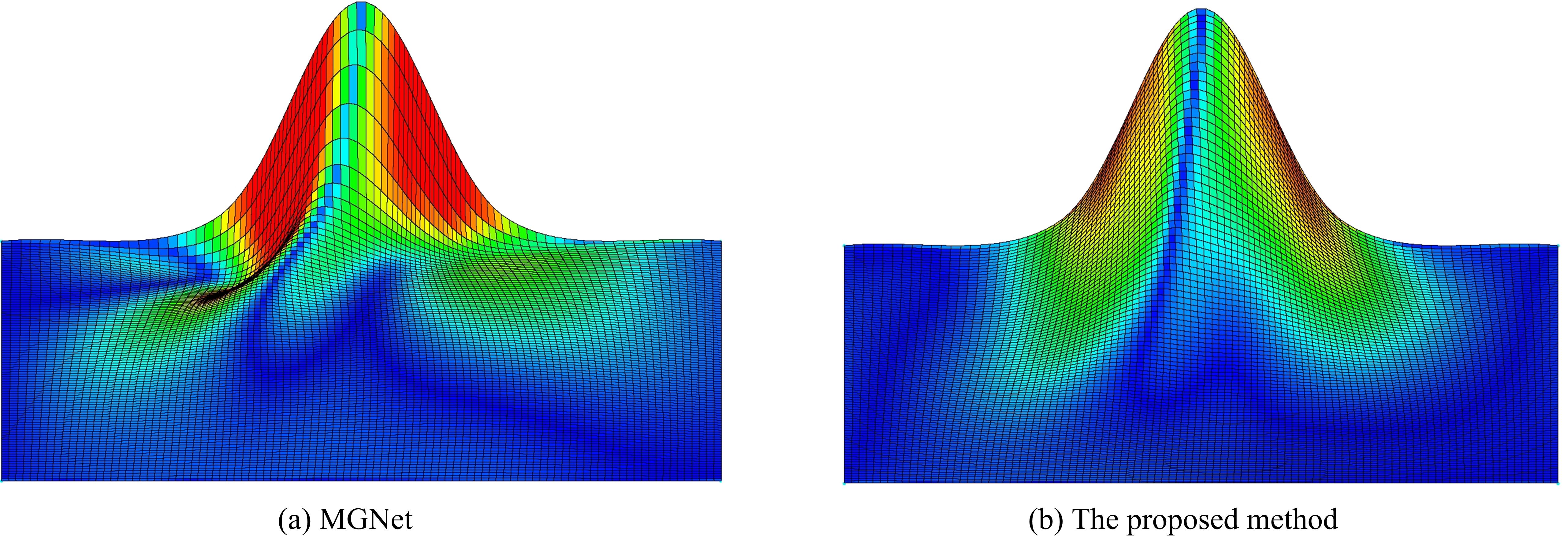}}
  \caption{The quality evaluation results of different neural network-based mesh generation methods.}
  \label{fig:fig6}
\end{figure}

Figure \ref{fig:fig5} visualizes the meshing results of two neural network-based generators governed by the Laplace equation. The results show that MGNet cannot always produce an acceptable mesh. The generated mesh suffers from cell degeneration inside the physical domain. These ‘sliver’ cells with poor orthogonality can negatively affect the overall quality of the obtained mesh and lead to inaccurate or non-convergence results during the simulation. In contrast, the proposed method offers more capable handling of 2-D structured meshing. When we input the points sampled on the auxiliary line, the underlying neural network is able to calibrate the subsequent optimization directions based on the values of these measured data. Benefiting from the introduced auxiliary line strategy, we can avoid the weaknesses of MGNet due to the suboptimal prediction. The trained network is capable of generating smooth and orthogonal mesh (see Figure \ref{fig:fig5}b).

In order to evaluate the quality of the generated meshes more intuitively, we employ $Maximum Included Angle$ as a quality metric to check the meshing results of different neural network-based methods. The results in Figure \ref{fig:fig6} demonstrate again that the proposed method can effectively improve the orthogonality of the generated mesh. The average included angle of the proposed method is 101.7 degrees, which is lower than that of MGNet (103.1 degrees). Meanwhile, the maximum included angle in the proposed method is 161.4 degrees, while 174 degrees for MGNet.

\begin{figure}[htbp]
  \centerline{\includegraphics[width=1\textwidth]{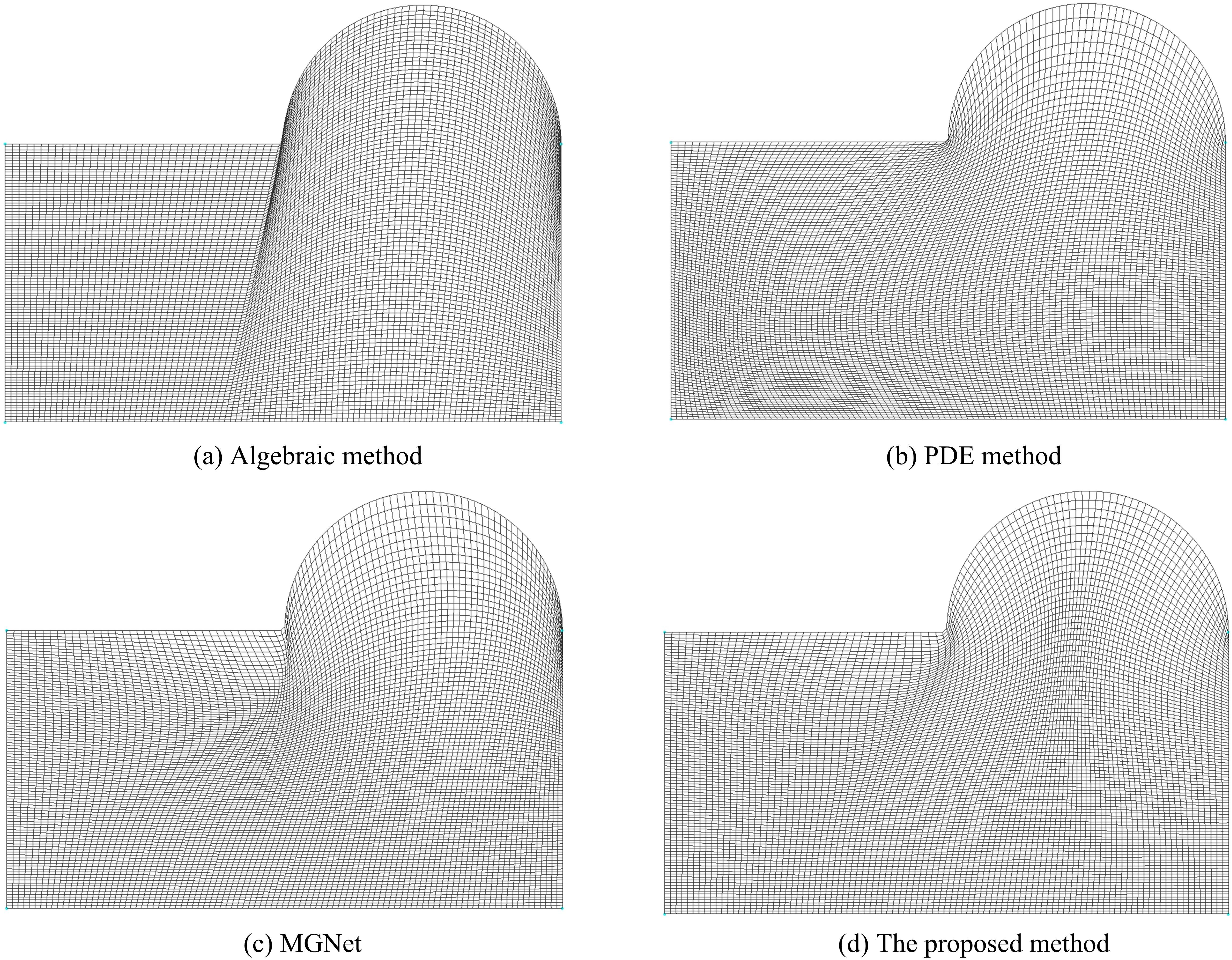}}
  \caption{Visualization results of different mesh generation methods.}
  \label{fig:fig7}
\end{figure}

\begin{figure}[htbp]
  \centerline{\includegraphics[width=1\textwidth]{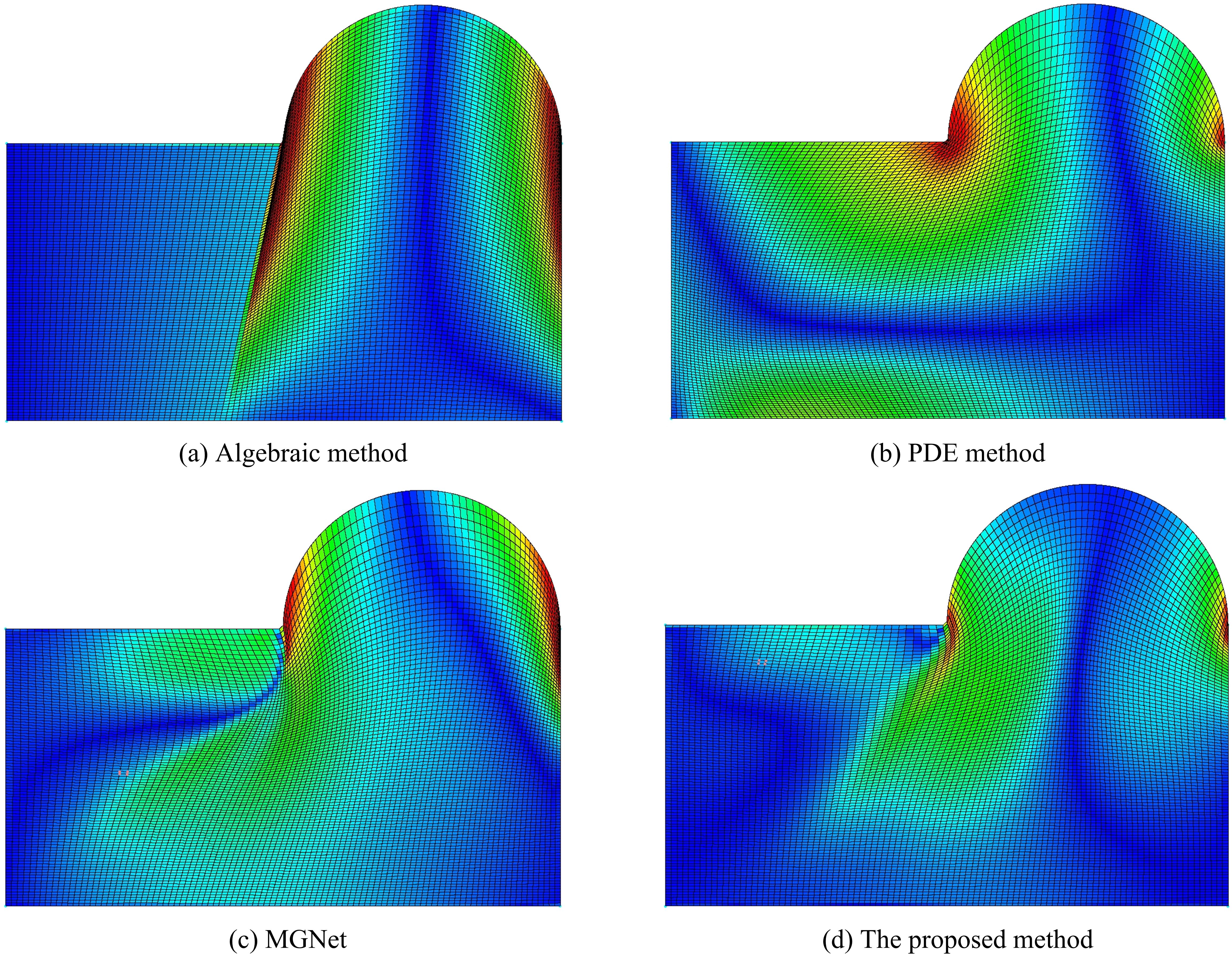}}
  \caption{The quality evaluation results of different mesh generation methods.}
  \label{fig:fig8}
\end{figure}

In the second test case, we use the geometry depicted in Figure \ref{fig:fig2} to evaluate the performance of the proposed method. For the sake of comparison, the results of algebraic and PDE methods are also shown in Figures \ref{fig:fig7} and \ref{fig:fig8}. The visualization results in these figures show that the algebraic method tends to generate poorly shaped cells in the unsmooth near-wall region. Due to insufficient training, MGNet can only slightly improve the orthogonality of the mesh cells near the boundary, but the overall quality of the mesh still needs to be optimized. For the PDE method, we initialize the coordinates to 0 and set the number of iterations to 1000. The results in Figure \ref{fig:fig7}b prove that, despite the expensive meshing overhead, the PDE method is able to generate meshes with relatively good orthogonality. However, it is clear that the proposed method allows an accurate approximation of the mapping from the computational domain to the physical domain and ultimately achieves comparable meshing results to the PDE method.

\begin{figure}[htbp]
  \centerline{\includegraphics[width=1\textwidth]{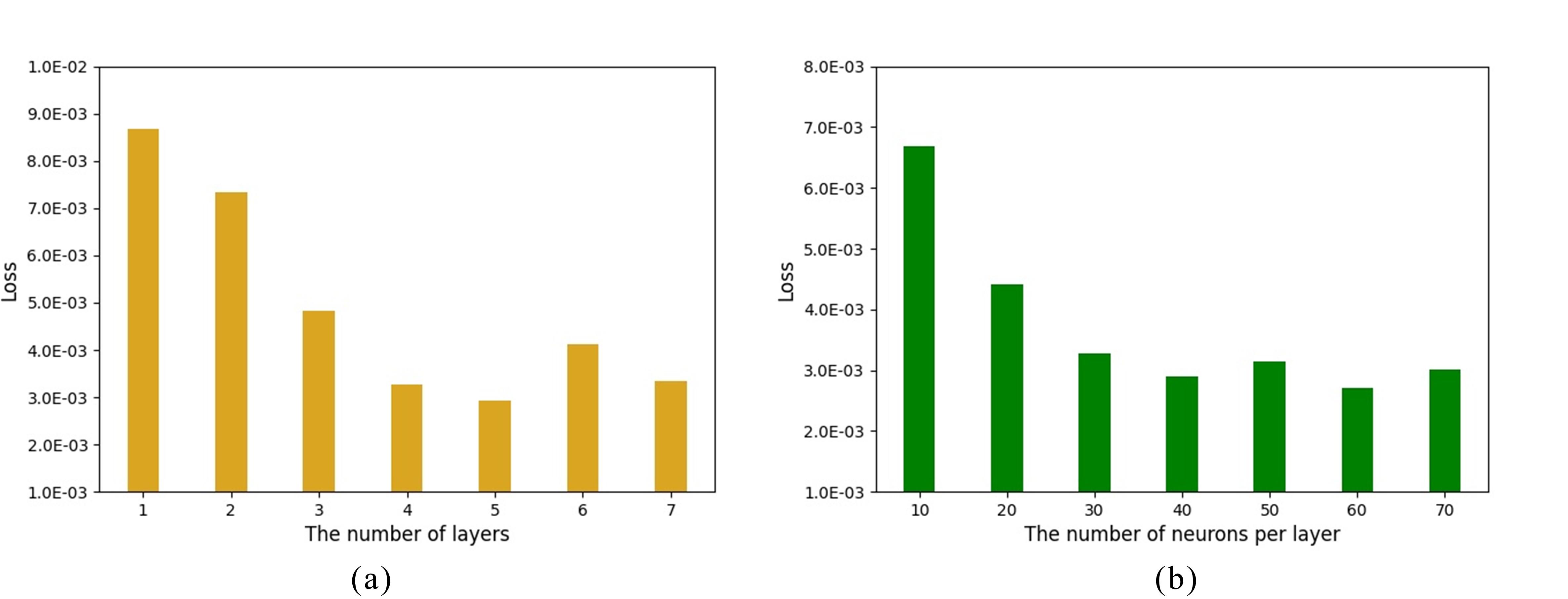}}
  \caption{Performances of different architectural designs obtained by varying the number of hidden layers (left) and the number of neurons per layer (right).}
  \label{fig:fig9}
\end{figure}

We also conduct experiments for different architectures, i.e., the number of hidden layers and the number of neurons per layer, to investigate their impact on the prediction solution. To simplify the comparison, we use the loss values to quantify the prediction performance. Figure \ref{fig:fig9}a illustrates the performance when the number of single-layer grid cells is 30, and the number of network layers is varied from 1 to 7. We can see that a single-layer network design tends to return relatively inaccurate predictions. By increasing the number of layers, the proposed method is able to obtain better approximation results. However, we can also observe that using deeper networks may not guarantee better performance. An excessive number of layers may lead to an undesirable network deformation, resulting in suboptimal results. A similar conclusion can be obtained in Figure \ref{fig:fig9}b. This figure analyzes the performance for different numbers of neurons per layer (the network layer is fixed at 4). It can be seen that an increase in the number of neurons does not necessarily improve the prediction performance. The network is relatively saturated when the number of neurons per layer is 30.

Overall, we present an improved physics-informed neural network method, which can be used as an efficient structured mesh generator. This new method uses the governing equation, the boundary constraints, and the measured data (sampled from the auxiliary line strategy) to establish a composite loss function. Subsequently, we treat the meshing problem as an optimization problem and find a solution that best fits the mapping from the computational domain to the physical domain using a well-designed neural network. The trained network is able to generate the mesh through the feedforward prediction technique, which enables fast and high-quality mesh generation. The resulting meshes are shown for several example geometries.

\section{Conclusion}
The automatic generation of computational meshes is one of the prerequisites for any attempt to perform high-resolution numerical simulations. In this paper, we develop an improved structured mesh generation method based on physics-informed neural networks. In contrast to traditional meshing methods, we formulate the mesh generation problem as a global optimization problem, more precisely, as an approximation problem to find a valid mapping from the computational domain to the physical domain. For this purpose, we propose a novel auxiliary line strategy and an efficient neural network model. Experimental results demonstrate that the proposed method is capable of generating smooth, orthogonal, high-quality meshes in different two-dimensional scenarios. An added benefit of our method is that the meshing overhead is low due to the efficient neural network feedforward prediction technique.

Although the current work is implemented for two-dimensional problems, we will actively pursue the extension to three-dimensional cases in future work. Moreover, the proposed method is basically a two-stage process. In the first stage, suitable auxiliary lines are selected to generate measured data as ground truth. The meshing process is then performed by a physics-informed neural network. While the auxiliary line strategy offers an efficient way to mitigate the misprediction or distortion in complex regions, this strategy is inherently empirical and might introduce extra human intervention. Thus, it is also interesting to investigate an efficient auxiliary line selection mechanism for the fully automatic meshing process.

\section*{Acknowledgments}
This research work was supported in part by the National Key Research and Development Program of China (2021YFB0300101).

\bibliographystyle{unsrt}  
\bibliography{ref-aux}

\end{document}